\documentclass[12pt,twocolumn,A4]{article}

\usepackage{graphics}
\begin{document}
\onecolumn
\baselineskip 0.25in
\title{\Huge{Persistent Current in Metallic Rings and Cylinders}}
\author{\Large {Santanu K. Maiti} \\ \\
        \Large {E-mail: {\em santanu.maiti@saha.ac.in}} \\ \\
        \Large {$^1$Theoretical Condensed Matter Physics Division} \\
        \Large {Saha Institute of Nuclear Physics} \\
        \Large {1/AF, Bidhannagar, Kolkata-700 064, India} \\ \\
        \Large {$^2$Department of Physics} \\
        \Large {Narasinha Dutt College} \\
        \Large {129, Belilious Road, Howrah-711 101, India}}
\date{}
\maketitle
\newpage
\tableofcontents

\newpage
\begin{center}
\addcontentsline{toc}{section}{\bf {Preface}}
{\Large \bf Preface}
\end{center}

\noindent
We explore the behavior of persistent current and low-field magnetic 
response in mesoscopic one-channel rings and multi-channel cylinders
within the tight-binding framework. We show that the characteristic
properties of persistent current strongly depend on total number of
electrons $N_e$, chemical potential $\mu$, randomness and total number
of channels. The study of low-field magnetic response reveals that only
for one-channel rings with fixed $N_e$, sign of the low-field currents 
can be predicted exactly, even in the presence of disorder. On the other 
hand, for multi-channel cylinders, sign of the low-field currents cannot
be mentioned exactly, even in the perfect systems with fixed $N_e$ as
it significantly depends on the choices of $N_e$, $\mu$, number of 
channels, disordered configurations, etc.

\vskip 0.2in
\begin{flushleft}
\end{flushleft}

\newpage
\section{Introduction}

The physics of small metallic rings provides an excellent testing ground
for many ideas of basic physics. In thermodynamic equilibrium, a small
metallic ring threaded by a magnetic flux $\phi$ supports a current that
does not decay dissipatively even at non-zero temperature. It is the
so-called persistent current in mesoscopic normal metal rings. This 
phenomenon is a purely quantum mechanical effect and gives an obvious 
demonstration of the Aharonov-Bohm effect~\cite{aharo}. The possibility 
of persistent current was predicted in the very early days of quantum 
mechanics by Hund~\cite{hund}, but their experimental evidences came much 
later only after realization of the mesoscopic systems. In $1983$, 
B\"{uttiker} {\em et al.}~\cite{butt} predicted that persistent current 
can exist in mesoscopic normal metal rings threaded by a magnetic flux 
even in the presence of impurity. Few years later, in a pioneering 
experiment Levy {\em et al.}~\cite{levy} first gave the experimental 
evidence of persistent current in mesoscopic metallic rings.
Following with this, the existence of the persistent current was further
confirmed by several experiments~\cite{mailly1,chand,jari,deb,reul,rab}. 
Though there exists a vast literature of theoretical~\cite{butt1,cheu1,cheu2,
land,byers,von,mont,bouc,san2,alts,schm,abra,mull,kulik,orella1,kulik2,san11,
san12,san14,san13} as well as experimental~\cite{levy,mailly1,chand,jari,
deb,reul,rab} results on persistent currents, but lot of controversies are 
still present between the theory and experiment. For our illustrations, 
here we mention very briefly some of them as follow. (i) The main 
controversy appears in the determination 
of the current amplitude. It has been observed that the measured current 
amplitude exceeds an order of magnitude than the theoretical estimates. 
Many efforts have been paid to solve this problem, but no such proper 
explanation has yet been found out. Since normal metals are intrinsically 
disordered, it was believed that electron-electron correlation can enhance 
the current amplitude by homogenize the system~\cite{giam}. But the 
inclusion of the 
electron correlation~\cite{san3} doesn't give any significant enhancement 
of the persistent current. Later, in some recent papers~\cite{san4,san10,
san5} it has been studied that the simplest nearest-neighbor
tight-binding model with electron-electron interaction cannot explain the
actual mechanisms. The higher order hopping integrals in addition to the
nearest-neighbor hopping integral have an important role to magnify the
current amplitude in a considerable amount. With this prediction some
discrepancies can be removed, but the complete mechanisms have to be
understood. (ii) The appearance of different flux-quantum periodicities
rather than simple $\phi_0$ ($\phi_0=ch/e$, the elementary flux-quantum)
periodicity in persistent current is not quite clear to us. The presence
of other flux-quantum periodicities has already been reported in many
papers~\cite{avishai,weiden,san8,san7}, but still there exist so many
conflict. (iii) The prediction of the sign of low-field currents is a 
major challenge in this area. Only for a single-channel ring, the sign 
of the low-field currents can be mentioned exactly~\cite{san7,san9}. 
While, in all other cases i.e., for multi-channel rings and cylinders, 
the sign of the low-field currents cannot be predicted properly. It then 
depends on the total number of electrons ($N_e$), chemical potential 
($\mu$), disordered configurations, etc. Beside these, there are several 
other controversies those are unsolved even today. Thus it can be 
emphasized that the study of persistent current in normal metal rings, 
cylinders and other loop geometries is an open challenge to us. In the
present article we investigate the behavior of persistent current in 
mesoscopic rings and cylinders within the non-interacting electron 
picture. This study may be helpful for the beginners to understand the 
basic features of persistent current in metallic loops.

The article is organized as follow. In Section $2$, we illustrate the
behavior of persistent current in mesoscopic normal metal rings and
cylinders. Section $3$ focuses the magnetic response of low-field 
currents both at absolute zero and finite temperatures. Finally, we 
conclude our results in Section $4$.

\section{Persistent Current in Non-Interacting Single-Channel and
Multi-Channel Mesoscopic Rings}

Our aim of this section is to study persistent currents in some small 
non-superconducting loops threaded by a magnetic flux $\phi$. A conducting 
ring, penetrated by a magnetic flux $\phi$, carries an equilibrium current 
in its ground state that {\em persists} in time. An 
electrically charged particle moving around the ring but not entering the 
region of magnetic flux, feels no (classical) force during its motion. 
However, the magnetic vector potential $\vec{A}$, related to the magnetic 
field through the relation $\vec{B}=\vec{\bigtriangledown} \times \vec{A}$, 
affects the quantum state of the particle by changing the phase of its wave 
function. As a consequence, both thermodynamic and kinetic properties 
oscillate with the magnetic flux $\phi$. Here we present some analytical 
as well as numerical calculations and study the behavior of persistent 
current $I$ in mesoscopic rings as a function of flux $\phi$, system size 
$L$, total number of electrons $N_e$, chemical potential $\mu$, strength 
of disorder $W$ and total number of channels.

\subsection{Origin of Persistent Current}

\begin{figure}[ht]
{\centering\resizebox*{5.25cm}{3cm}{\includegraphics{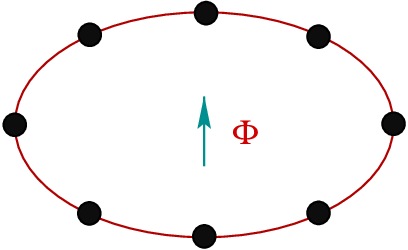}}
\par}
\caption{One-dimensional mesoscopic ring threaded by a magnetic flux $\phi$,
where the filled circles correspond to the position of the atomic sites. 
A persistent current $I$ is established in the ring.}
\label{figure1}
\end{figure}

In this sub-section we describe how persistent current appears in a small
normal metal ring threaded by a magnetic flux $\phi$. The schematic
representation of the system is given in Fig.~\ref{figure1}. The electric
field $\mathcal E$ associated with the magnetic field $B$ in the ring can
be expressed through the relation (Faraday's law),
\begin{equation}
\vec{\bigtriangledown}\times \vec{\mathcal E}=-\frac{1}{c}\frac{\partial
\vec{B}}{\partial t}
\label{equ1}
\end{equation}
Using the above relation we can determine the electric field $\mathcal E$
from the following expressions:
\begin{equation}
\oint_{S}(\vec{\bigtriangledown}\times \vec{\mathcal E}).\vec{dS}=
-\frac{1}{c}\oint_{S}\frac{\partial\vec{B}}{\partial t}.\vec{dS}=-\frac
{1}{c}\frac{\partial}{\partial t}\oint_{S}\vec{B}.\vec{dS}=-\frac{1}{c}
\frac{\partial\phi}{\partial t}
\label{equ2}
\end{equation}
where $S$ is the area enclosed by the ring. From the Stoke's theorem we
can write,
\begin{equation}
\oint_{loop}\vec{\mathcal E}.\vec{dl}=-\frac{1}{c}\frac{\partial \phi}
{\partial t}
\label{equ3}
\end{equation}
and thus the electric field can be written in terms of the time variation of
the magnetic flux $\phi$ as,
\begin{equation}
\mathcal E = - \frac{1}{2\pi r c}\frac{\partial \phi}{\partial t}
\label{equ4}
\end{equation}
where $r$ is the radius of the ring. Therefore, the force acting on an
electron in the ring becomes,
\begin{equation}
F=-\frac{e}{2\pi r c}\frac{\partial \phi}{\partial t}
\label{equ5}
\end{equation}
and the change in energy or work done for a small displacement
$\bigtriangleup s$ is,
\begin{equation}
\bigtriangleup E=\bigtriangleup W=\vec{F}.\vec{\bigtriangleup s}=
-\frac{e}{2\pi r c}\frac{\bigtriangleup \phi}{\bigtriangleup t}
\bigtriangleup s=-\frac{e}{2\pi r c}\bigtriangleup \phi\left(\frac
{\bigtriangleup s}{\bigtriangleup t}\right)
\label{equ6}
\end{equation}
The velocity of the electron in the ring can be expressed as,
\begin{equation}
v = \frac{\bigtriangleup s}{\bigtriangleup t}= -\frac{2\pi r c}{e}
\left(\frac{\bigtriangleup E}{\bigtriangleup \phi} \right)
\label{equ7}
\end{equation}
and the persistent current that developed in the ring becomes,
\begin{equation}
I=ef=\frac{ev}{2\pi r}=-c\left(\frac{\bigtriangleup E}{\bigtriangleup \phi}
\right)
\label{equ8}
\end{equation}
This is the final expression of persistent current, and we see that the
current carried by an energy eigenstate is obtained by taking the first 
order derivative of the energy for that particular state with respect to 
the magnetic flux $\phi$. For the sake of simplicity, throughout our 
studies we use the units where $c=1$, $e=1$ and $h=1$.

\subsection{Non-Interacting One-Channel Rings}

Here we focus our attention on the behavior of persistent currents in
one-channel rings~\cite{san9}, where all the currents are computed for the
non-interacting electron picture based on the tight-binding formulation. 
The model Hamiltonian for a $N$-site ring ($L=Na$, $a$ is the lattice 
spacing) threaded by a magnetic flux $\phi$ (in units of the elementary 
flux-quantum $\phi_0=ch/e$) can be expressed in this form,
\begin{equation}
H = \sum_{i}\epsilon_{i}c_i^{\dagger}c_i + \sum_{<ij>} t \left[e^{i\theta}
c_i^{\dagger} c_j + e^{-i\theta} c_j^{\dagger} c_i \right]
\label{equ9}
\end{equation}
where $c_i^{\dagger}$ ($c_i$) corresponds to the creation (annihilation)
operator of an electron at the site $i$, $t$ represents the nearest-neighbor
hopping strength, $\epsilon_i$'s are the on-site energies and $\theta=2 \pi
\phi/N$ is the phase factor due to the flux $\phi$ threaded by the ring.
The magnetic flux $\phi$ enters explicitly into the above Hamiltonian
(Eq.~\ref{equ9}), and the wave functions satisfy the periodic boundary
condition which is equivalent to consider the above Hamiltonian at zero
flux with the flux-modified boundary conditions:
\begin{eqnarray}
\left .\psi\right|_{x=L} & = & exp\left[\frac{2\pi i\phi}{\phi_{0}}\right]
\left .\psi\right|_{x=0} \nonumber \\
\left .\frac{d\psi}{dx}\right|_{x=L} & = & exp\left[\frac{2\pi i\phi}{\phi_0}
\right]\left .\frac{d\psi}{dx}\right|_{x=0}
\label{equ10}
\end{eqnarray}
here $x$ varies between $0$ to $L$ and is expressed as
$x=L\theta^{\prime}/2\pi$, where $\theta^{\prime}$ is the azimuthal angle,
the spatial degrees of freedom of the electron in the ring.

\subsubsection{Impurity Free Rings}

In order to reveal the basic properties of persistent currents, let us
begin our discussion with the simplest possible system which is the case
of impurity free non-interacting electron model.

\vskip 0.4cm
\noindent
{\bf Energy Spectra}
\vskip 0.2cm
\noindent
For a perfect ring, setting $\epsilon_i=0$ for all $i$, we get the energy 
\begin{figure}[ht]
{\centering\resizebox*{10.0cm}{6cm}{\includegraphics{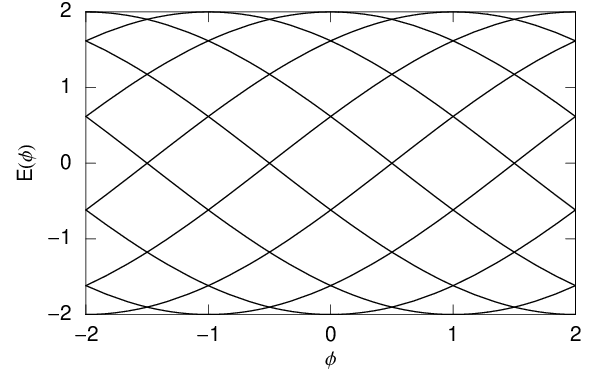}}\par}
\caption{Electron energy levels as a function of the magnetic flux $\phi$
for a one-dimensional impurity-free ring considering the system size $N=10$.}
\label{figure2}
\end{figure}
of $n$th eigenstate as,
\begin{equation}
E_{n}(\phi)=2t \cos\left[\frac{2\pi}{N}\left(n+\frac{\phi}{\phi_{0}}\right)
\right]
\label{equ11}
\end{equation}
where $n=0,\pm1,\pm2,\ldots$. In Fig.~\ref{figure2}, we plot the energy-flux
($E$-$\phi$) characteristics for a typical ordered ring considering its size
$N=10$. From the spectrum it is observed that the energy levels of the ring 
vary periodically with $\phi$ showing $\phi_0$ flux-quantum periodicity.
At an integer or half-integer flux quantum, the energy levels have an
extrema i.e., either a maximum or a minimum, and accordingly, at these 
values of $\phi$ persistent current should vanish since it is computed from
the first order derivative of the energy eigenvalue. In the following, we 
discuss the current-flux ($I$-$\phi$) characteristics for the two different 
cases. For one case we consider the rings with fixed number of electrons 
$N_e$, while for the other case the rings have some fixed chemical potential 
$\mu$.

\vskip 0.4cm
\noindent
{\bf Persistent Current: Rings with Fixed $N_e$}
\vskip 0.2cm
\noindent
The current carried by the $n$th energy eigenstate, whose energy is given by
Eq.~\ref{equ11}, can be obtained through the expression,
\begin{equation}
I_{n}(\phi)=\left(\frac{4\pi t}{N\phi_{0}}\right)\sin\left[\frac{2\pi}{N}
\left(n+\frac{\phi}{\phi_{0}}\right)\right]
\label{equ12}
\end{equation}
At absolute zero temperature ($T=0$ K), the total persistent current is 
obtained
\begin{figure}[ht]
{\centering\resizebox*{9cm}{10cm}{\includegraphics{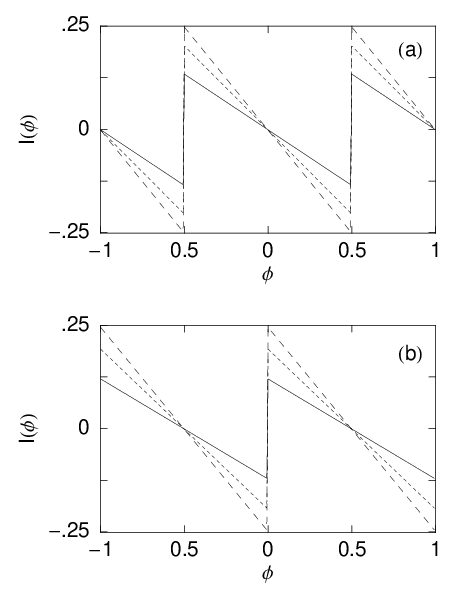}}\par}
\caption{Current-flux characteristics for some one-channel perfect rings
($N=50$) described with fixed number of electrons. The solid, dotted and
dashed curves in (a) correspond to $N_e=9$, $15$ and $23$ electrons 
respectively, while in (b) these curves correspond to $N_e=8$, $14$ and 
$22$ electrons respectively.}
\label{figure3}
\end{figure}
by taking the sum of individual contributions from the lowest $N_e$ energy
eigenstates. For our illustrative purpose, in Fig.~\ref{figure3}, we display 
the current-flux characteristics for some typical one-channel perfect rings 
($N=50$), where (a) corresponds to the results for the rings with odd $N_e$ 
and (b) represents the results for the rings with even $N_e$. The results
predict that the current shows saw-tooth like nature with sharp transitions 
at half-integer and integer flux quanta for the rings with odd and even 
$N_e$ respectively. For all such cases, the current varies periodically 
with $\phi$ providing $\phi_0$ flux-quantum periodicity.

\vskip 0.4cm
\noindent
{\bf Persistent Current: Rings with Fixed $\mu$}
\vskip 0.2cm
\noindent
For the rings described with fixed chemical potential $\mu$, instead of 
$N_e$, the total persistent current at $T=0$ K will be obtained by adding 
all the individual contributions from the energy levels
\begin{figure}[ht]
{\centering\resizebox*{10cm}{10cm}{\includegraphics{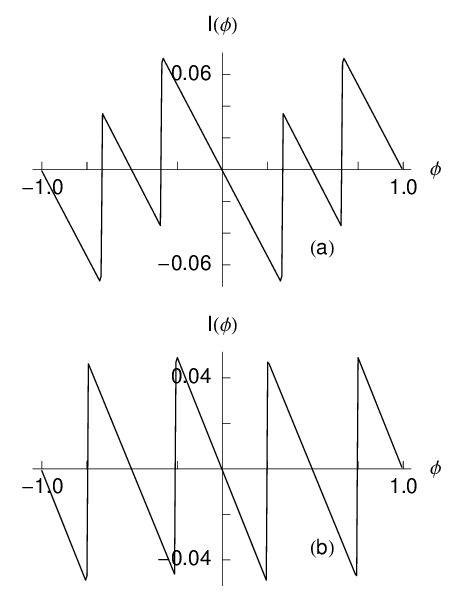}}\par}
\caption{Current-flux characteristics for the one-channel perfect rings
($N=100$) described with constant $\mu$, where (a) $\mu=-1$ and
(b) $\mu=-1.25$.}
\label{figure4}
\end{figure}
with energies less than or equal to $\mu$. As the chemical potential is 
fixed, the total number of electrons varies as a function of the magnetic 
flux $\phi$ except for some special choices of $\mu$, where the rings 
contain fixed number of electrons. In Fig.~\ref{figure4}, we plot the 
persistent currents for some one-channel perfect rings those are described 
with fixed chemical potential $\mu$. Here we take the ring size $N=100$. 
Figures~\ref{figure4}(a) and (b) correspond to the persistent currents for 
the rings with $\mu=-1$ and $-1.5$ respectively. Our results predict that 
several additional kink-like structures appear at different field points 
and their positions also depend on the choices of $\mu$. For all these 
cases the current gets only $\phi_0$ periodicity, as expected.

\subsubsection{Rings with Impurity}

Metals are intrinsically disordered which tends to decrease persistent
current due to the localization effect~\cite{lee1} of energy eigenstates.
In order to emphasize the role of impurities on persistent currents now
we concentrate our study on the rings in the presence of disorder.

\vskip 0.4cm
\noindent
{\bf Energy Spectra}
\vskip 0.2cm
\noindent
To introduce the impurities in the ring, we choose the site energies 
$\epsilon_i$'s randomly from a ``Box" distribution function of width 
$W=1$, which reveal that the ring is subjected to the diagonal disorder. 
In the presence of impurity in the ring, gaps open at the points 
of intersection
\begin{figure}[ht]
{\centering\resizebox*{10.0cm}{6cm}{\includegraphics{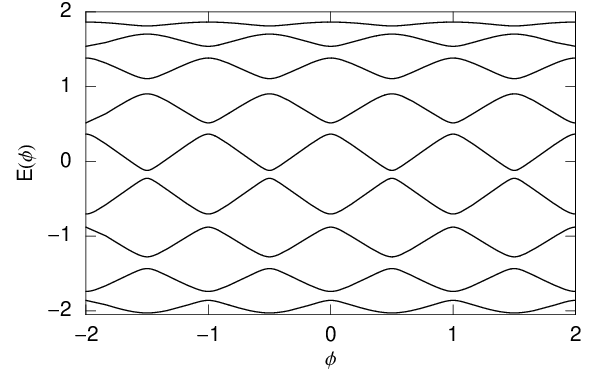}}
\par}
\caption{Electron energy levels as a function of the flux $\phi$ for
a one-channel ring ($N=10$) in the presence of impurity with strength
$W=1$.}
\label{figure5}
\end{figure}
of the energy levels, in the same way as band gaps form in the 
band-structure problem, and they vary continuously with the magnetic flux 
$\phi$. Figure~\ref{figure5} shows the variation of the energy eigenstates 
as a function of $\phi$ for a one-channel ring ($N=10$) in the 
presence of diagonal disorder. The continuous variation of the energy 
levels with respect to $\phi$ is due to the removal of the degeneracies 
of the energy eigenstates in the presence of impurity in the ring.

\begin{figure}[ht]
{\centering\resizebox*{9cm}{10cm}{\includegraphics{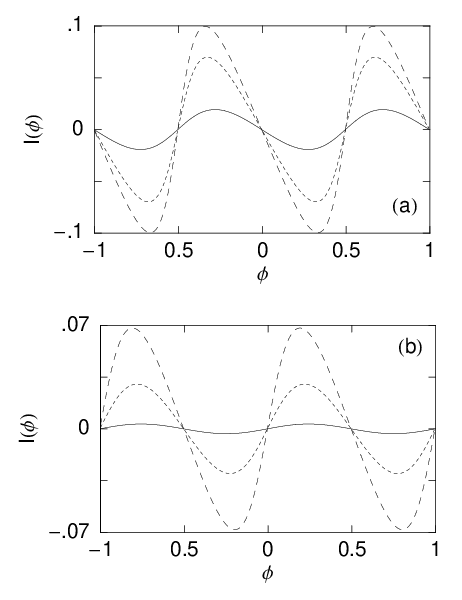}}\par}
\caption{$I$-$\phi$ characteristics for the disordered ($W=1$) one-channel
rings ($N=50$) with fixed number of electrons. The solid, dotted and
dashed curves in (a) correspond to $N_e=9$, $15$ and $23$ electrons 
respectively, while in (b) they correspond to $N_e=8$, $14$ and $22$ 
electrons respectively.}
\label{figure6}
\end{figure}

\vskip 0.4cm
\noindent
{\bf Persistent Current: Rings with Fixed $N_e$}
\vskip 0.2cm
\noindent
In Fig.~\ref{figure6}, we show the variation of the persistent currents for 
some one-channel disordered rings considering the ring size $N=50$ and the 
impurity strength $W=1$. Figure~\ref{figure6}(a) corresponds to the 
persistent currents for the
rings described with odd number of electrons, while Fig.~\ref{figure6}(b)
represents the currents for the rings with even number of electrons. It is
observed that the current varies continuously as a function of $\phi$ and
gets much reduced amplitude compared to the results obtained in the
impurity-free rings (see Fig.~\ref{figure3}). The continuous variation of
the current is clearly visible from the variation of the energy spectrum
since they become continuous as long as the impurities are introduced in
the ring. On the other hand, the suppression of the current amplitudes
is due to the localization effect~\cite{lee1} of the energy eigenstates 
in the presence of impurity. Here all the results are described for some 
typical disordered configurations of the ring, and in fact we examine that 
the qualitative behavior of the persistent currents do not depend on the 
specific realization of the disordered configurations. This is the generic 
feature of persistent current for any one-channel non-interacting rings 
in the presence of impurity those are described with fixed number of 
electrons $N_e$.

\vskip 0.4cm
\noindent
{\bf Persistent Current: Rings with Fixed $\mu$}
\vskip 0.2cm
\noindent
The behavior of the current-flux characteristics is quite interesting in
\begin{figure}[ht]
{\centering\resizebox*{9cm}{9cm}{\includegraphics{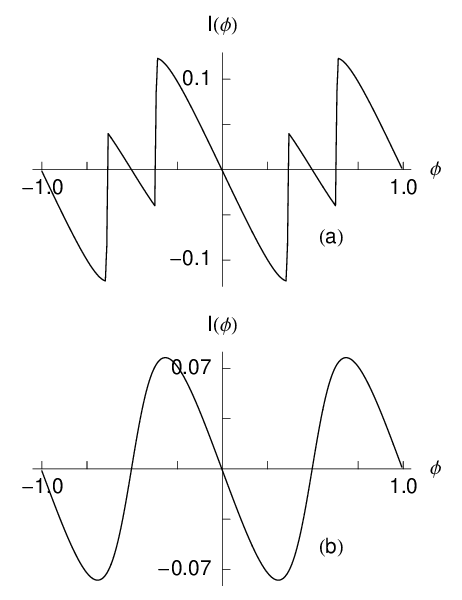}}\par}
\caption{$I$-$\phi$ characteristics for the disordered ($W=1$) one-channel
rings ($N=50$) described with constant $\mu$, where (a) $\mu=-1$ and
(b) $\mu=-1.25$.}
\label{figure7}
\end{figure}
the presence of impurity for the rings described with fixed chemical 
potential $\mu$, instead of $N_e$. As representative
example, in Fig.~\ref{figure7} we plot the $I$-$\phi$ characteristics for
some one-channel disordered rings those are described with fixed $\mu$, 
where (a) represents the result for the ring with $\mu=-1$ and (b) denotes 
the result for the ring considering $\mu=-1.5$. From these results we can 
emphasize that depending on the choices of $\mu$ the current shows different 
behavior as a function of $\phi$ and in all such cases the current exhibits 
only $\phi_0$ periodicity.

\subsection{Non-Interacting Multi-Channel Systems}

\begin{figure}[ht]
{\centering\resizebox*{6.0cm}{4cm}{\includegraphics{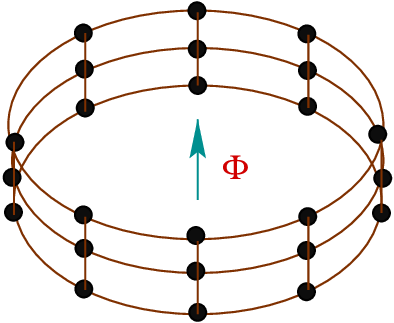}}
\par}
\caption{Schematic view of a multi-channel cylinder threaded by a magnetic
flux $\phi$.}
\label{figure8}
\end{figure}

Now we focus our study on the behavior of persistent currents in
non-interacting multi-channel systems~\cite{san9}. A schematic representation
of such a multi-channel system of cylindrical geometry threaded by a magnetic
flux $\phi$ is given in Fig.~\ref{figure8}.
Considering the lattice spacing both in the longitudinal and transverse
directions are identical i.e., the surface of the cylinder forms a square
lattice, we can write the Hamiltonian of the system by the tight-binding
formulation as,
\begin{equation}
H = \sum_x \epsilon_x c_x^{\dagger}c_x + \sum_{<xx^{\prime}>}\left[
t_{x x^{\prime}} e^{i\theta_{x x^{\prime}}} c_x^{\dagger} c_{x^{\prime}}
+ t_{x x^{\prime}} e^{-i\theta_{x x^{\prime}}} c_{x^{\prime}}^{\dagger} c_x
\right]
\label{equ21-1}
\end{equation}
where $\epsilon_x$ is the site energy of the lattice point $x$ of coordinate,
say, ($i,j$). $t_{xx^{\prime}}$ is the hopping strength between the lattice
points $x$ and $x^{\prime}$ and $\theta_{xx^{\prime}}$ is the phase factor
acquired by the electron due to the longitudinal hopping in the presence of
magnetic flux $\phi$. The study of persistent currents in such multi-channel
systems becomes much more relevant compared to strictly one-dimensional rings
(see Fig.~\ref{figure1}), where we get only one channel that carries current,
since most of the conventional experiments are performed in rings with finite
width. Here we will describe the characteristic properties of persistent
currents for some non-interacting multi-channel rings concerning the
dependence of the current on total number of electrons $N_e$, chemical
potential $\mu$, strength of disorder $W$ and number of channels. All the
results studied here are performed only at absolute zero temperature.

\subsubsection{Energy Spectra}

\begin{figure}[ht]
{\centering\resizebox*{10cm}{11cm}{\includegraphics{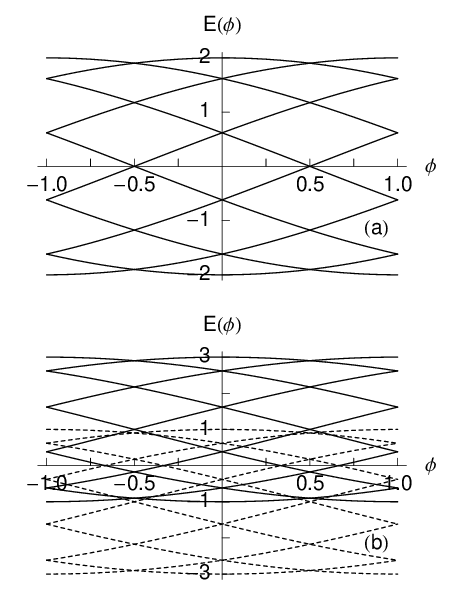}}\par}
\caption{Energy levels as a function of the magnetic flux $\phi$, for
(a) the perfect ring (one layer) with $N=10$ and (b) the perfect cylinder
with two layers taking $N=10$ in each of these layers.}
\label{figure9}
\end{figure}

In order to present the behavior of persistent currents in mesoscopic
multi-channel systems, let us first describe the energy-flux characteristics
of a small cylindrical system that threads a magnetic flux $\phi$.
To have a deeper insight to the problem we begin our discussion with
the simplest possible system which can be calculated analytically.
This is the case of a two-layer impurity free cylinder threaded by a magnetic
flux $\phi$. This cylindrical system can be treated as two one-channel rings
placed one above the other and they are connected by some vertical bonds
(like as in Fig.~\ref{figure8}). For strictly one-dimensional ring i.e.,
for one layer the energy of $n$th eigenstate is expressed in the form
$E_n(\phi) = 2 t\cos\left[\frac{2\pi}{N}\left(n+\frac{\phi}{\phi_0}\right)
\right]$ (see Eq.~\ref{equ11}), where $n=0$, $\pm 1$, $\pm 2$, $\ldots$.
Here $t$ is the nearest-neighbor hopping strength and $N$ is the total
number of lattice points in the ring. The behavior of such energy levels
as a function of flux $\phi$ is shown in Fig.~\ref{figure9}(a), where we
take $N=10$. From this figure it is observed that the energy levels are
bounded within the range $-2$ to $2$ in the scale of $t$ and the crossing
of the energy levels (flux points where the energy levels have degeneracy)
occurs at half-integer or integer multiples of $\phi_0$. Now as we add
another one ring with the previous one-dimensional ring and connect it by
$N$ vertical bonds it becomes a cylinder with two layers. For such a system,
we get two different energy bands of discrete energy levels, each of which
contains $N$ number of energy levels and they are respectively expressed
in the form,
$E_{1n}(\phi)=t+2t\cos\left[\frac{2\pi}{N}\left(n+\frac{\phi}{\phi_0}
\right)\right]$ and
$E_{2n}(\phi)=-t+2t\cos\left[\frac{2\pi}{N}\left(n+\frac{\phi}{\phi_0}
\right)\right]$, where the symbol $n$ corresponds to the same meaning as
above and $t$ is the nearest-neighbor hopping strength which is identical
both for the longitudinal and transverse directions. In Fig.~\ref{figure9}(b),
we plot the energy levels of a small impurity-free cylinder considering
$N=10$, where the solid and dotted curves correspond to the energy levels
in the two separate energy bands respectively. These two different energy
bands are bounded respectively in
the range $-1$ to $3$ and $-3$ to $1$ in the scale of $t$. Accordingly, an
overlap energy region appears for the two energy bands in the range $-1$ to
$1$ as shown in the figure (Fig.~\ref{figure9}(b)). In this overlap region,
the energy levels cross each other at several other flux points in addition
to the half-integer and integer multiples of $\phi_0$ which provide different
characteristic features of persistent currents. For cylinders with more
than two layers, we get more energy bands like above and therefore several
other overlap energy regions appear in energy spectra.

In the presence of impurity in multi-channel cylinders, gaps open at the
crossing points of the energy levels and they become a continuous function
of the flux $\phi$, like as in the one-channel disordered rings
(Fig.~\ref{figure5}). In all such perfect and disordered multi-channel
cylinders, energy levels vary periodically with period $\phi_0$.

Now we shall describe the characteristic features of persistent currents
for some small cylindrical systems, and our results might be quite helpful
to explain the characteristic properties of persistent currents
for larger system sizes.

\subsubsection{Persistent Current}

\begin{figure}[ht]
{\centering\resizebox*{12cm}{13cm}{\includegraphics{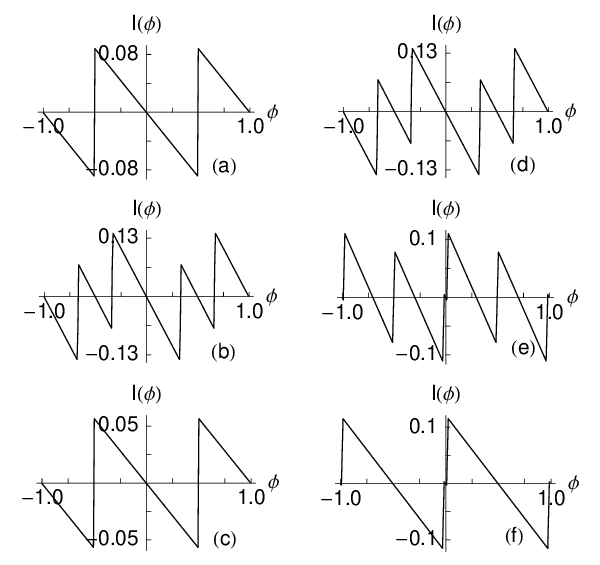}}
\par}
\caption{$I$ versus $\phi$ curves of the perfect cylinders with two layers
taking $N=100$ in each layer. The results for the fixed $N_e$ are plotted in
the first column, where (a) $N_e=25$, (b) $N_e=100$ and (c) $N_e=185$, while
in the second column the results are plotted for the fixed $\mu$, where
(d) $\mu=0$, (e) $\mu=-0.5$ and (f) $\mu=-1.5$.}
\label{figure10}
\end{figure}

\vskip 0.3cm
\noindent
{\bf Perfect Cylinders}
\vskip 0.2cm
\noindent
In this sub-section we study persistent currents of some impurity-free
multi-channel systems of cylindrical geometry with two layers concerning
the dependence of the current on total number of electrons $N_e$ and
chemical potential $\mu$. As illustrative example, in Fig.~\ref{figure10}
we plot the current-flux characteristics for some perfect cylinders
considering $N=100$ in each of these two layers. Let us first describe
the behavior of persistent currents for the cylinders those are described
with fixed number of electrons. The first column of Fig.~\ref{figure10}
corresponds to the currents for the systems with fixed $N_e$. To emphasize
the effect of the energy overlap region on persistent currents here we study 
the systems for three different values of $N_e$. Figures~\ref{figure10}(a)
and (c) correspond to the currents for the systems with $N_e=25$ (low) and 
$N_e=185$ (high) respectively. The result for the intermediate value of 
$N_e$ ($N_e=100$) is shown in Fig.~\ref{figure10}(b). It is shown that both 
for the low and high values of $N_e$, the persistent currents get saw-tooth 
like nature with sharp transitions only at half-integer multiple of $\phi_0$, 
similar to that of strictly
one-channel impurity-free rings with odd $N_e$ (see Fig.~\ref{figure3}(a)).
This behavior can be explained as follow. For the system with $N_e=25$,
the net persistent current is obtained by taking the sum of the lowest $25$
energy eigenstates those lie below the energy overlap region. Now away from
this overlap region, the energy levels behave exactly in the same way with
strictly one-channel rings and therefore the current shows similar kind of
saw-tooth shape as observed in one-channel rings. On the other hand, for the
system with $N_e=185$ the situation is quite different than the previous
one. To obtain the net current for this case we cross the overlap energy
region since the highest energy level that contributes current lies far
above of this overlap region. The net contribution to the current from the
energy levels within this overlap region vanishes and therefore no new
feature appears in the persistent current compared to the system with
$N_e=25$. Now we describe the result plotted in Fig.~\ref{figure10}(b),
where the system contains intermediate value of $N_e$ ($N_e=100$). For
such a case the persistent current shows some additional kink-like
structures across $\phi=\pm 0.5$. These kinks are due to the different 
contributions of the energy levels in the overlap energy region, since 
in the overlap region energy levels have more degeneracy at several 
other flux points rather than the half-integer and integer multiples of 
$\phi_0$. For other multi-channel systems with more than two layers, 
some more kinks may appear in persistent current at different values 
of $\phi$ depending on the choice of $N_e$ and the number of layers.

In multi-channel systems where we fix the chemical potential $\mu$
instead of the total number of electrons $N_e$,
we also get different kink-like structures in the persistent currents as
observed from the results plotted in the second column of Fig.~\ref{figure10}.
For all such systems described either with fixed $N_e$ or $\mu$, the current
exhibits only $\phi_0$ flux-quantum periodicity.

\vskip 0.4cm
\noindent
{\bf Dirty Cylinders}
\vskip 0.2cm
\noindent
To illustrate the effect of the impurity on persistent currents in
multi-channel systems, here we concentrate our study in some dirty
\begin{figure}[ht]
{\centering\resizebox*{12.0cm}{8cm}{\includegraphics{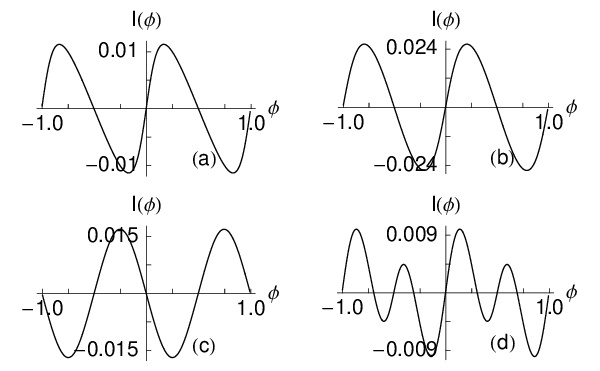}}
\par}
\caption{Current-flux characteristics for the disordered ($W=1$) cylinders
with two layers considering $N=100$ in each layer. The results for the fixed
$N_e$ are plotted in the first column, where (a) $N_e=50$ and (b) $N_e=100$,
while the results for the constant $\mu$ are plotted in the second column,
where (c) $\mu=0$ and (d) $\mu=-0.5$.}
\label{figure11}
\end{figure}
cylinders considering the same system size as taken above for the
perfect systems. In Fig.~\ref{figure11}, we plot the current-flux
characteristics for some disordered cylinders where the impurities
are given only in the site energies ($\epsilon_x$) by choosing
them randomly from a ``Box" distribution function of width $W=1$. All the
results are plotted for isolated disordered configurations of the system.
The first column of Fig.~\ref{figure11} represents the persistent currents
for the cylinders described with fixed $N_e$,
while the second column of this figure corresponds to the currents for the
systems with fixed $\mu$. It is observed that, in all these cases the
persistent current varies continuously as a function of flux $\phi$
showing $\phi_0$ periodicity. The explanation of such a continuous
variation has already been described in our previous studies. Another 
significant point observed from Figs.~\ref{figure11}(a) and (b) is that, 
the slope of the current in the zero-field
limit ($\phi \rightarrow 0$) changes in opposite direction though both for
the two cases the systems contain even number of electrons. This provides
the signature of random sign of low-field currents in multi-channel 
systems. The detailed description of low-field magnetic response
will be available in the forthcoming section (Section $3$).

Thus we can emphasize that the behavior of persistent current in
multi-channel systems strongly depends on the disordered configurations,
total number of electrons $N_e$, chemical potential $\mu$ and also on the
total number of channels.

\section{Low-Field Magnetic Response on Persistent Current}

The diamagnetic or the paramagnetic sign of low-field persistent currents 
also becomes a
controversial issue due to discrepancy between theory and experiment. From
the theoretical calculations, Cheung {\em et al.}~\cite{cheu2} predicted
that the sign of persistent current is random depending on the total number
of electrons, $N_e$, in the system and on the specific realization of the
disordered configurations of the ring. Both the diamagnetic and the
paramagnetic responses were also observed theoretically in mesoscopic Hubbard
rings by Yu and Fowler~\cite{yu}. They showed that the rings with odd $N_e$
exhibit the paramagnetic response, while those with even $N_e$ have the
diamagnetic response in the limit $\phi \rightarrow 0$. In other recent 
work Waintal {\em et al.}~\cite{wain} have also discussed theoretically 
the sign of persistent current of $N$ electrons in one-dimensional rings 
and provided several interesting results. In an experiment on $10^7$ 
isolated mesoscopic Cu rings, Levy {\em et al.}~\cite{levy} had reported 
the diamagnetic response 
for the low-field currents, while with Ag rings Chandrasekhar 
{\em et al.}~\cite{chand} got the paramagnetic response. In a recent 
experiment, Jariwala {\em et al.}~\cite{jari} have got the diamagnetic
persistent currents with both integer and half-integer flux-quantum
periodicities in an array of $30$-diffusive mesoscopic gold rings. The
diamagnetic sign of the currents in the vicinity of zero magnetic field were
also found in an experiment~\cite{deb} on $10^5$ disconnected Ag ring. The
sign of the low-field current is a priori not consistent with the theoretical
predictions. In this section, we will study the nature of the low-field
magnetic susceptibility of mesoscopic rings and cylinders through some 
exact calculations.

The magnetic susceptibility of any mesoscopic ring/cylinder can be obtained
from the general expression~\cite{san9},
\begin{equation}
\chi(\phi)=\frac{N^3}{16\pi^2}\left[\frac{\partial I(\phi)}{\partial \phi}
\right]
\label{equ24}
\end{equation}
Calculating the sign of $\chi(\phi)$, one can predict whether the current
is paramagnetic or diamagnetic. Here we focus our attention on the systems
either with fixed number of electrons ($N_e$) or with fixed chemical
potential ($\mu$).

\subsection{One-Channel Mesoscopic Rings}

\begin{figure}[ht]
{\centering \resizebox*{10cm}{10cm}{\includegraphics{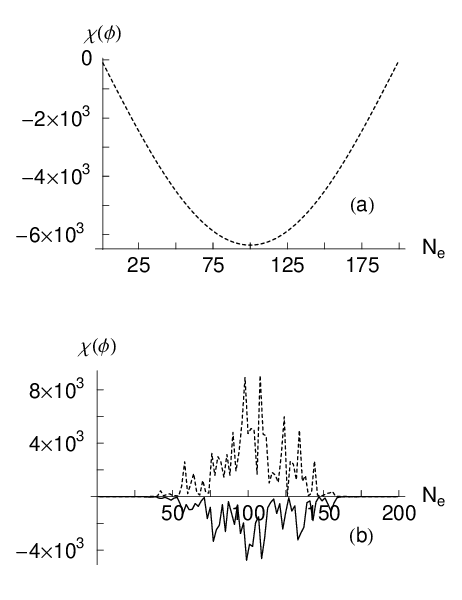}}\par}
\caption{Low-field magnetic susceptibility as a function of $N_e$, for
(a) perfect rings ($W=0$) and (b) rings in the presence of impurity ($W=1$)
with $N=200$. The solid and dotted lines in (b) correspond to the rings 
with odd and even $N_e$ respectively.}
\label{figure35}
\end{figure}

Let us first study the low-field magnetic susceptibility of impurity-free
one-channel mesoscopic rings with fixed $N_e$. Figure~\ref{figure35}(a) shows
the variation of $\chi(\phi)$ as a function of $N_e$ for a perfect ring with
$N=200$ in the limit $\phi \rightarrow 0$. It is noticed that, both for the
even and odd $N_e$ the current has only the diamagnetic sign. This diamagnetic
sign of the low-field currents can be easily understood from the slope of the
current-flux characteristics of the one-channel impurity-free rings (see
the curves plotted in Fig.~\ref{figure3}). From these curves it follows that
the persistent current always exhibits negative slope at low-fields. Therefore,
it can be predicted that for perfect one-channel rings the current shows only
the diamagnetic sign near zero-field limit, irrespective of the total number
of electrons $N_e$ i.e., whether the rings contain odd or even $N_e$.

The effects of disorder on the low-field currents are quite interesting, and
our results illustrate that the sign of the currents, even in the presence of
disorder, can be mentioned without any ambiguity both for the rings with odd
and even $N_e$. In Fig.~\ref{figure35}(b), we plot $\chi(\phi)$ as a function
of $N_e$ for the disordered rings. Here we take $N=200$ and $W=1$. The solid
and dotted lines in Fig.~\ref{figure35}(b) correspond to the results for
the rings with odd and even $N_e$ respectively. These curves show that the
rings with odd $N_e$ exhibit only the diamagnetic sign for the low-field
currents, while for even $N_e$ the low-field currents always have the
paramagnetic sign. Physically, the disorder lifts all the degeneracies of the
energy levels those were observed in a perfect ring, and as a result the sharp
discontinuities of the $I$-$\phi$ characteristics (see the curves of
Fig.~\ref{figure3}) disappear. It may be noted that the slopes of the
$I$-$\phi$ curves for even and odd $N_e$ always have opposite signs near
zero magnetic field (see the curves of Fig.~\ref{figure6}). Thus for the
one-dimensional disordered rings with fixed number of electrons, the sign of
the low-field current is independent of the specific realization of the
disordered configurations and depends only on the oddness or evenness of
$N_e$.

\subsubsection{Effect of Temperature}

At finite temperature, we notice an interesting behavior of the low-field
magnetic susceptibility of mesoscopic rings.
\begin{figure}[ht]
{\centering\resizebox*{9cm}{6cm}{\includegraphics{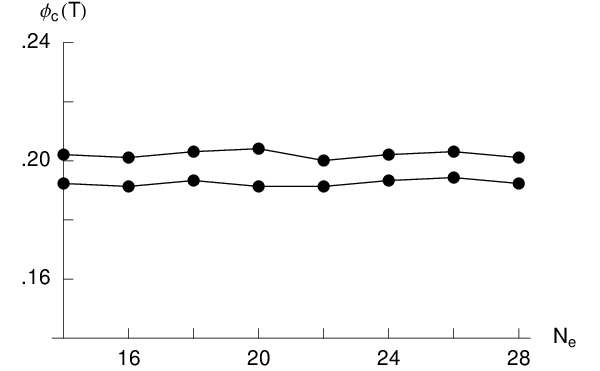}}\par}
\caption{$\phi_c(T)$ versus $N_e$ ($N_e$=even) curves for perfect rings with
size $N=45$.}
\label{figure36}
\end{figure}
Let us confine ourselves to the systems with even number of electrons. At any
finite temperature, the magnetic response of these systems are always
paramagnetic both for the perfect and the dirty rings in the zero field limit.
For a given system, this paramagnetism is observed over a certain range of
$\phi$ close to $\phi=0$, say, in the domain $-\phi_0/4 \leq \phi \leq
\phi_0/4$. Quite interestingly we observe that, at finite temperature the
magnetic response of this particular system becomes diamagnetic beyond a
critical field $\phi_c(T)$, even though $|\phi_c(T)| < \phi_0/4$.

\begin{figure}[ht]
{\centering \resizebox*{9cm}{6cm}{\includegraphics{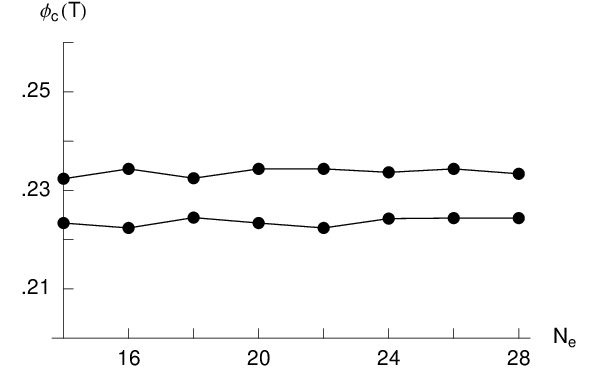}}\par}
\caption{$\phi_c(T)$ versus $N_e$ ($N_e$=even) curves for dirty ($W=1$)
rings with size $N=40$.}
\label{figure37}
\end{figure}

In Fig.~\ref{figure36}, we show the variation of the critical field
$\phi_c(T)$ with respect to only even $N_e$ for a perfect one-channel ring
of size $N=45$. The curve with higher values of $\phi_c(T)$ corresponds
to the temperature $T/T^{\star}=1.0$, while the other curve corresponds to 
$T/T^{\star}=0.5$. Figure~\ref{figure37} represents the behavior
of $\phi_c(T)$ for a dirty sample (with $W=1$) at the same two temperatures
$T/T^{\star}=1.0$ (upper curve) and $T/T^{\star}=0.5$ (lower curve). From
Figs.~\ref{figure36} and \ref{figure37} it is clear that the critical
value of $\phi$, where the transition from the paramagnetic to the diamagnetic
phase takes place, increases with the increase of the temperature. Thus we see
that, both for the perfect and dirty rings with even number of electrons
there exists a critical value of magnetic flux $\phi_c(T)$, beyond which the
magnetic response of the low-field currents exhibits a transition from the
paramagnetic to the diamagnetic phase.

The situation is quite different even at zero temperature when we describe
the system by constant chemical potential instead
of fixed $N_e$. It may be noted that, only for some particular values
of $\mu$ the system will have a fixed number of electrons, and for these
values of $\mu$ the sign of the low-field currents can be predicted according
to the above prescriptions. While, for all other choices of $\mu$ the total
number of electrons varies even for a slight change in magnetic flux $\phi$
in the neighborhood of zero flux. Hence, it is not possible to predict the
sign of the low-field currents precisely, even in the absence of any impurity
in the system. Thus the sign of the low-field currents strongly depends on
the choice of $\mu$, the strength of disorder and the choice of the
disordered configurations.

\subsection{Multi-Channel Mesoscopic Cylinders}

We have also studied the low-field magnetic response for the mesoscopic
rings of finite width~\cite{san9}. Our study reveals that, for such
systems it is not possible to predict the sign of the low-field currents
precisely even for the impurity-free cases with fixed number of electrons.
So we can conclude that,
in the diffusive multi-channel mesoscopic rings the sign of the low-field
currents is a highly unpredictable quantity as it can be easily affected by
the total number of electrons $N_e$, chemical potential $\mu$, magnetic 
flux $\phi$, strength of disorder $W$, realizations of
disordered configurations, etc. This is exactly the same picture what has
been observed experimentally regarding the sign of the low-field currents.

\section{Concluding Remarks}

In this dissertation, we have demonstrated the behavior of the persistent 
current and low-field magnetic response in mesoscopic one-channel rings
and multi-channel cylinders. All the results have been computed in the 
non-interacting electron picture withing the tight-binding framework.
These results may be quite helpful for the beginners to realize the basic
features of persistent current in metallic loops.

The characteristic properties of persistent current in the non-interacting 
one-channel rings and multi-channel cylinders have been presented in Section
$2$ showing its dependence on the total number of
electrons $N_e$, chemical potential $\mu$, randomness and the total number
of channels. All the calculations have been performed only at absolute zero
temperature. In perfect one-channel rings with fixed $N_e$, persistent
current shows saw-tooth like behavior as a function of magnetic flux $\phi$ 
with sharp discontinuities at $\phi=\pm n\phi_0/2$ or $\pm n \phi_0$ 
depending on whether the system has odd or even $N_e$. On the other hand, 
some additional kinks may appear in the currents for the one-channel perfect 
rings with fixed $\mu$. The situation is quite different for the 
multi-channel perfect cylinders. In such cylindrical rings, the kinks appear 
in the persistent currents for both the cases with fixed $N_e$ or fixed $\mu$.

The diamagnetic or the paramagnetic sign of the low-field currents is a
controversial issue due to the discrepancy between theory and experiment. 
In Section $3$, we have examined the behavior of the low-field magnetic 
response of persistent currents by calculating the magnetic susceptibility 
in the limit $\phi \rightarrow 0$. In perfect one-channel rings, the 
low-field current
exhibits only the diamagnetic sign irrespective of the parity of the
total number of electrons $N_e$ i.e, whether $N_e$ is odd or even, while
in the disordered rings currents have the diamagnetic or the paramagnetic
nature depending on whether the rings contain odd or even $N_e$. The
important point is that, for the disordered one-channel rings with
fixed $N_e$ the sign of the low-field currents is completely independent of
the specific realization of the disordered configurations. In this context
we have also studied the effect of finite temperature and observed that
both for the perfect and the dirty rings with even number of electrons,
there exits a critical value of magnetic flux $\phi_c(T)$ beyond which the
magnetic response of the low-field currents makes a transition from the
paramagnetic to the diamagnetic phase. But in dirty rings with constant 
chemical potential $\mu$, the sign of the low-field currents cannot be 
predicted since it strongly depends on the choices of $\mu$. Finally, in 
the case of multi-channel systems we have noticed that the sign of these 
currents cannot be predicted exactly, even in the perfect systems with 
fixed $N_e$ as it strongly depends on the choice of $N_e$, $\mu$, number 
of channels, disordered configurations, etc.

\addcontentsline{toc}{section}{\bf {References}}

\end{document}